\documentclass[12pt,a4paper]{article}
\usepackage[a4paper,margin=25mm]{geometry}
\usepackage[T1]{fontenc}
\usepackage[utf8]{inputenc}
\usepackage{graphicx}
\usepackage{textcomp}
\usepackage{ulem}
\usepackage{amsmath}
\usepackage{amssymb}
\usepackage{xcolor}
\usepackage{framed}
\usepackage{pifont}
\usepackage{upgreek}
\usepackage{times}
\usepackage{bm}
\usepackage{tensor}
\usepackage{empheq}
\usepackage[makeroom]{cancel}
\usepackage{needspace}
\usepackage{xspace}
\usepackage{relsize}
\usepackage{verbatim}
\usepackage{float}
\usepackage{subfig}
\usepackage{nicefrac}

\newcommand{\im}{\mathrm{i}}
\newcommand{\am}{\mathrm{a}}

\newcommand{\dd}{\mathrm{d}}
\newcommand{\e}{\mathrm{e}}
\newcommand{\Lag}{L}

\newcommand{\Hf}{H}

\newcommand{\HCd}{\mathcal{\Hf}}

\newcommand{\LCd}{\mathcal{\Lag}}

\newcommand{\QCd}{\mathcal{Q}}
\newcommand{\NCd}{\mathcal{N}}
\newcommand{\PCd}{\mathcal{P}}
\newcommand{\OCd}{\mathcal{O}}

\newcommand{\rmi}{\im}
\newcommand{\IM}{\mathrm{Im}}

\renewcommand{\d}{\,\mathrm{d}}

\newcommand{\dt}{\d{t}}
\newcommand{\dtau}{\d{\tau}}
\newcommand{\onehalf}{{\textstyle\frac{1}{2}}}

\newcommand{\quarter}{{\textstyle\frac{1}{4}}}

\newcommand{\ihalf}{{\textstyle\frac{\rmi}{2}}}

\newcommand{\phat}{\hat{p}}
\newcommand{\ahat}{\hat{a}}
\newcommand{\Hhat}{\hat{\HCd}}
\newcommand{\bra}{\langle}
\newcommand{\ket}{\rangle}
\newcommand{\braU}{\langle U|}
\newcommand{\Uket}{|U \rangle}
\newcommand{\Ukett}{|U \rangle|_\tau}
\newcommand{\dota}{\dot{a}}

\newcommand{\pfrac}[2]{\frac{\partial{#1}}{\partial{#2}}}

\newcommand{\sref}[1]{Section~\ref{#1}}

\newcommand{\eref}[1]{Eq.~(\ref{#1})}

\newcommand{\keywords}[1]{\par\noindent\textbf{Keywords:} #1}

\title{On measuring the Quantum Universe}
\author{
David Vasak$^{1}$\thanks{Correspondence: \texttt{vasak@fias.uni-frankfurt.de}}, Johannes Kirsch$^{1}$\\
J\"urgen Struckmeier$^{1,2}$\\[0.75em]
\small $^{1}$Frankfurt Institute for Advanced Studies (FIAS), Ruth-Moufang-Str.\ 1, 60438 Frankfurt am Main, Germany\\
\small $^{2}$Goethe Universit\"at, Max-von-Laue-Str.\ 1, 60438 Frankfurt am Main, Germany
}
\date{}

\begin{document}

\bibliographystyle{unsrt}

\maketitle

\begin{abstract}
We present a theoretical analysis of the WDW approach to quantum cosmology extended to gravity theories with torsion.
The dynamics of the FLRW universe is formulated as a classical Hamiltonian problem of point particle mechanics.
Unlike in the WDW formalism, the Hamiltonian is not zero, though, and the 3rd quantization does not enforce the cosmic time to vanish. 
The wave function of the Universe appears as a superposition of eigenfunctions of the quantum Hamiltonian with the cosmic time being the conjugate to its eigenvalues, spatial curvatures.
The notion of \emph{weak measurement} is then introduced to avoid the collapse of the total universal wave function upon measurements of the parameter set describing matter and spacetime.
The collapse postulate of the standard Copenhagen quantum theory is discussed and the de Broglie-Bohm interpretation of the effective wave function introduced.  
The question of the boundary conditions for both, the wave function and the Bohmian guidance equation, is addressed.
The corresponding numerical calculations will be published in a separate paper.
\\
\textit{Received: 25 March 2026. Revised: 25 March 2026. Accepted: 25 March 2026.}
\end{abstract}


\section{Introduction} \label{sec:Intro}
By the paradigm developed over the past century, everything \emph{in} this world is quantum. 
Can then the world \emph{itself}, the Universe, be exempt from this law? 
If not, how should its consistent description as a quantum system be constructed? 
How, then, can we interpret astronomical observations in view of the quantum mechanical axiom of the measurement process?
And, last not least, how does the classical Universe that we see around us emerge from the quantum domain?

\medskip
The most popular quantum version of standard cosmology has been developed by Wheeler and DeWitt \cite{Wheeler:1957mu, DeWitt1967} more than 60 years ago. 
Their "WDW" approach is based on the reduction of the full field-theoretical Einstein-Hilbert action to a subspace of the geometry, called \emph{mini-superspace}. 
The generic metric in the action integral is thereby replaced by the Friedman metric with two time-dependent parameters, the scale and lapse functions\footnote{The FLRW curvature parameter is usually dropped assuming the space geometry to be flat.}.
Since the corresponding Hamiltonian is found to be constrained to zero,  
its quantized version with the scale and the conjugate momentum considered  quantum operators, leads to a  Schr\"{o}dinger equation with zero eigenvalues.
The resulting wave function is thus static, and the time coordinate has simply disappeared.
The evolution of the Universe is frozen. 

\medskip
The motivation for the following study is to analyze how a quite generic formulations of the theory of gravity beyond General Relativity (e.g.\ with torsion), and an alternative of the WDW quantization process, could impact the quantum model 
of the Universe, and shed new light on the role of time\footnote{See also the critique on the WDW formulation and an independent way to recover the cosmic time as the evolution parameter in~\cite{Vasak:2025mjo}}.
The obvious absence of the collapse of the wave function of the Universe after astronomical observation is explained with the concept of "weak"(or sub-quantum) measurements.
And the fact that the astronomers and other star gazers are not external observers but constitute a part of the Universe, and that they observe as a classical system, is addressed by the Bohm-de Broglie "pilot wave" interpretation of quantum mechanics.

\medskip
The paper is organized as follows:

\medskip

In Section~\ref{sec:CCGG} the Concordance model of the Universe is revisited and the extended Friedman equations derived accommodating the gauge gravity ansatz. 

Section \ref{sec:QMA} is then devoted to the canonical quantization procedure and the derivation of the Schr\"odinger equation for the wave function of the Universe and the discussion of the related  quantum mechanics in configuration space. 
Rather than following the renown WDW  \emph{{a} priori} approach to 
mini-superspace quantization,  the canonical quantization is carried out \emph{{a} posteriori}, i.e.\  first, before quantization, a particular formulation of the Friedman equation is selected. 
The classical Hamiltonian is \emph{then} subject to the canonical quantization process (3$^{rd}$ quantization as classified e.g. by Kucha\v{r} in~\cite{Kuchar:1991qf}). 
The ambiguities with respect to the operator ordering are avoided, and we also retain the  curvature parameter $K$ of the Friedman metric in the framework. 
It indeed turns out to be crucial for the existence of a universal time.  
The important question of the boundary conditions of the wave function is addressed. 
We also review the polar (eikonal) representation of the wave function, the significance of the so called quantum potential, and its impact on the equations of motion.

\medskip
The preparation of the wave function of the Universe in alignment with existing astronomical observations relies on the concept of the so called  
"weak measurements".
This enables collecting information about the background wave function without triggering its collapse, and that information is used to restrict the physical Hilbert space and the wave function to its ``effective'' form. 

\medskip
For a consistent interpretation of the universal wave function we furthermore resort in Section \ref{sec:debrogliebohm} to the de Broglie-Bohm pilot-wave model of quantum mechanics. 
This, in contrast to the orthodox Copenhagen framework, allows to resolve the problem of missing external observers. 
Bohm's guidance equation, its non-local nature, and the significance of the initial conditions for the extraction of a classical evolution of the background Universe are discussed.  

\medskip
The summary and an outlook on future work are given in Section \ref{sec:summary}.

\section{Modelling the Universe} \label{sec:CCGG}
\subsection{The Concordance model}\label{sec:concordance}

The (semi-)classical model of the Universe considered in the following is based on the so called Cosmological Principle postulating isotropic and homogeneous distribution of matter across space. 
This constraints the geometry of space-time to a so called ``mini-superspace'' expressed by the choice of the FLRW (for Friedman-Lema\^{\i}tre-Robertson-Walker) metric  with two parameters, the dynamical dimensionless scale factor, ${a(t)}$ of the  3D-space expanding in (the universal) time $t$,
and the spatial constant curvature parameter $K$ of dimension~$L^{-2}$ (in natural units $\hbar = c = 0$):
\begin{align} \label{def:FLRW}
	ds^2 = dt^2-a^2(t) \left[\frac{dr^2}{1-K r^2}+
	r^2(d\theta ^2+\sin^2\theta d\phi ^2) \right].
\end{align}
$K$ is a continuous constant defining the spatial curvature of the mini-superspace (spherical, flat or hyperbolic) for~$K>0,\,K=0,\,K<0$ respectively. 
With this metric the Levi-Civita portion of the affine connection and the Riemann curvature can be calculated and then fed into the field equation.

\medskip
The matter content of the Universe is thereby approximated by ideal ``co-moving'', non-interacting fluids with the densities~$\rho_i(t)$ and pressures~$p_i(t)$ for baryonic and dark matter~($i=\,$m), and for radiation~($i=\,$r), with the barotropic equations of state (EOS):
\begin{equation} \label{EOSgeneric}
	p_i = \omega_i\, \rho_i.
\end{equation}
For General Relativity with a phenomenological cosmological constant~$\Lambda$ this gives the so called $\Lambda$CDM or Concordance model: 
Inserting Eqs.~\eqref{def:FLRW} and \eqref{EOSgeneric} in Einstein's field equation, 
\begin{align}\label{CCGG}
R^{(\mu\nu)}-\tfrac{1}{2}g^{\mu\nu}R -\Lambda g^{\mu\nu} = 
8\pi G\, T_\text{matter}^{(\mu\nu)},
\end{align}
leads to the well known Friedman-Lema\^{\i}tre equations for $a(t)$:
\begin{subequations}
	\begin{align}
		{\left(\frac{\dot{a}}{a}\right)}^2+\frac{K}{a^2}-\frac{1}{3}\Lambda &= \frac{8\pi G}{3} \sum_{i=m,r} \rho_i
		\label{eq:f1a} \\
	    \frac{\ddot{a}}{a} - \frac{1}{3}\Lambda  &= -\frac{4\pi G}{3} \sum_{i=m,r} \left(\rho_i + 3p_i\right).
		\label{eq:f2a}
	\end{align}
\end{subequations}
Here $G$ is Newton's gravitational constant and the dot denotes time derivatives with respect to the universal cosmic time $t$, i.e.\ $\dot{a} \equiv \d a /\d t$.
Under the assumptions, the fluids be inert and non-interacting, the combination of equations~\eqref{eq:f1a}, \eqref{eq:f2a} and \eqref{EOSgeneric} fixes the dynamics of each individual fluid:
\begin{equation}
	\dot{\rho}_i = -3\frac{\dot{a}}{a}\left(\rho_i + p_i  \right)
	=-3\frac{\dot{a}}{a}\,\rho_i \left(1 + \omega_i  \right).
\end{equation}
The EOS parameter for pressure-less matter is $\omega_\mathrm{m} = 0$, and for radiation we get $\omega_\mathrm{r} =  \nicefrac{1}{3}$.
The dynamical impact of the spatial curvature and of the cosmological constant emerges thereby too in the form of geometric ``fluids'' with fixed equations of state, namely with $\omega_\mathrm{K} = -\nicefrac{1}{3}$ and
$\omega_\Lambda = -1$.
With the definition of the so called Hubble parameter,
\begin{equation}
	H(a) := \frac{\dot{a}}{a}\,, \label{eq:Hubblefunc}
\end{equation}
the first Friedman equation \eqref{eq:f1a} can then be re-written as
\begin{equation}
 H^2(a) = \frac{8\pi G}{3} \rho = \frac{8\pi G}{3} \sum_{i=m,r,K,\Lambda} \rho_i\,.		\label{eq:FriedmanGR}
\end{equation}
In the following the time coordinate $t$ is replaced for convenience by the dimensionless time parameter $\tau$ with
\begin{subequations}\label{eq:densities}
\begin{align}
	\tau :&= H_{100}\,t \label{normaltime}\\
	a(t(\tau) ) &\equiv a(\tau).
\end{align}
$H_{100}$ is a universal reference value setting the scale of the theory (constant lapse), with the 
overdot denoting henceforth the derivative $\dd / \dd \tau$.
The individual contributions $\rho_i$ to the total energy density, $\rho_{\mathrm{}}$, are re-expressed by 
the cosmological parameters  $\Omega_i,\,i=\,$ m,\,r,\,$\Lambda$,\,K which are constant dimensionless relative densities aligned with the conventions of the $\Lambda$CDM model:
	\begin{align}
		\rho_\mathrm{m} &\equiv \rho_{\mathrm{crit}}^{100}\,\hat{\rho}_\mathrm{m}  := \rho_{\mathrm{crit}}^{100}\,\Omega_\mathrm{m} \, a^{-3} \label{eq:rm} \\
		\rho_\mathrm{r} &\equiv \rho_{\mathrm{crit}}^{100}\,\hat{\rho}_\mathrm{r}  := \rho_{\mathrm{crit}}^{100}\, \Omega_\mathrm{r} \, a^{-4} \label{eq:rr} \\
		\rho_\Lambda &\equiv \rho_{\mathrm{crit}}^{100}\,\hat{\rho}_\mathrm{\Lambda}  := \rho_{\mathrm{crit}}^{100}\, \Omega_{\Lambda} \label{eq:rl} \\
		\rho_\mathrm{K} &\equiv \rho_{\mathrm{crit}}^{100}\,\hat{\rho}_\mathrm{K}  := \rho_{\mathrm{crit}}^{100}\, \Omega_\mathrm{K} \, a^{-2} \label{eq:rK}
	\end{align}
with 
\begin{align}
   H_{100} &:= 100 \frac{km}{s\ Mpc} \\
 \rho_{\mathrm{crit}}^{100} &:= \frac{3H_{100}^2}{8\pi G} \label{eq:rc} \\
		\Omega_{\Lambda} &:=  \frac{\Lambda}{3H_{100}^2} \label{eq:Ol} \\
		\Omega_\mathrm{K} &:= -\frac{K}{H_{100}^2}\,. \label{eq:Ok}
	\end{align}
\end{subequations}
Notice that with $a_0 = a(\tau_0)$ denoting the present-day scale of the Universe, $H_0 = H(a_0) =:h H_{100}$ stands for the present-day value of the Hubble parameter, hence $h~\approx 0.64 - 0.72$.
\eref{eq:FriedmanGR} is then equivalent to
\begin{equation}
	E^2(a) = \frac{H^2(a)}{H_{100}^2} 
	= 
    \hat{\rho}_\mathrm{K}+\hat{\rho}_{\Lambda}+ \hat{\rho}_\mathrm{{m}} + \hat{\rho}_\mathrm{{r}} \, ,
	\label{eq:HubbleGR}
\end{equation}
where 
\begin{equation}\label{def:E(tau)}
    E(a) := \frac{1}{a}\,\frac{\dd a}{\dd \tau} 
    = \frac{1}{a}\,\frac{\dd a}{\dd t}\,\frac{\dd t}{\dd \tau} = \frac{H(a)}{H_{100}}
\end{equation}
is the normalized Hubble function.
Setting w.l.o.g. the scale $a_0$ of the present universe (at time $\tau = \tau_0$) to 1, $a(\tau_0)=a_0=1$, gives $E(1) = h$, and thus
\begin{align}
	 h^2 = \Omega_\mathrm{m} +\Omega_\mathrm{r} +\Omega_{\Lambda} +\Omega_\mathrm{K} \, . \label{eq:sumOm}
\end{align}

\subsection{The Concordance model revisited}\label{sec:concordancerev}

The Friedman equation~\eqref{eq:HubbleGR} of the $\Lambda$CDM model can be interpreted as the vanishing sum of the energy-momentum tensors of matter and spacetime - see the discussion of the Zero-energy condition e.g.\ in Refs. \cite{lorentz1916, levi-civita1917, Tryon:1973xi, He:2014yia, Vasak:2023iaz}.
This interpretation suggest to express it as
\begin{equation}\label{eq:energybalancerev}
 \rho_{\mathrm{crit}}\left(\hat{\rho}_\mathrm{st} + \hat{\rho}_\mathrm{matter}\right) = 0,
\end{equation}
such that the partial energy densities of matter and spacetime must always cancel. 
Since we assume for physical reasons that the energy density of matter is always positive, the contribution of the energy density attributed to the dynamics and geometry of spacetime must be negative:
\begin{subequations}
  \begin{align}
    \hat{\rho}_\mathrm{matter}  :=& \,\hat{\rho}_\mathrm{r} + \hat{\rho}_\mathrm{m} \ge 0 \\
    \hat{\rho}_\mathrm{st} :=& \,\hat{\rho}_\mathrm{kin} +\hat{\rho}_\mathrm{K}+\hat{\rho}_{\Lambda} \le 0.
    \end{align}
\end{subequations}
The individual energy density contributions as they appear in~\eref{eq:HubbleGR} can obviously be complemented by $\hat{\rho}_\mathrm{kin} := -E^2(a)$.
While the leading negative term is the kinetic energy of spacetime given by the Hubble term, the energy densities associated with spatial curvature and the cosmological constant depend on the value and sign of the parameters K and $\Lambda$.

\medskip
We assume in the following that the above densities 
will be derived from some (extended) theory of gravity adding a term $\hat{\rho}_\mathrm{ext}(a)$ to~\eref{eq:HubbleGR}\footnote{Examples for $\hat{\rho}_\mathrm{{ext}}$ can be contributions from higher-order curvature terms, torsion of spacetime, and/or non-metricity, see e.g. \cite{Kirsch:2023iwd,vandeVenn:2022gvl}.}:
\begin{equation}
	E^2(a) = 
	 \hat{\rho}_\mathrm{K}+\hat{\rho}_{\Lambda} + \hat{\rho}_\mathrm{{m}} + \hat{\rho}_\mathrm{{r}}+\hat{\rho}_\mathrm{{ext}} .
	\label{eq:HubbleGRextgen}
\end{equation}
The sum of the density terms on the r.h.s. will in general be approximated by a Laurent polynomial or Laurent series in the scale parameter $a$.


\section{The quantum mechanical approach} \label{sec:QMA}

\subsection{Hamiltonian formulation}
We expect that even in the extended version of gravity the constant $\Omega_\mathrm{K}$ retains the form \eqref{eq:Ok}, and the remaining densities depend on a set of parameters that we generically call $\Omega_i$  or simply $\Omega$, but do not depend on the spatial curvature parameter $K$, i.e.\ $i \ne K$.
Upon multiplying~\eref{eq:HubbleGRextgen} by $a^2$, a short calculation shows its equivalence to
\begin{subequations}\label{eq:Fr12}
\begin{align}
		&\dota^2 + V(a;\Omega) = \Omega_\mathrm{K}\label{eq:Fr12a} \\ 
    &V(a;\Omega) := -a^2\left(
    \hat{\rho}_{\Lambda}+\hat{\rho}_\mathrm{{ext}} + \hat{\rho}_\mathrm{{m}} + \hat{\rho}_\mathrm{{r}}\right).\label{eq:Fr12pot}
\end{align}
The dependence of $V(a)$ on the "Concordance" parameters $\Omega_i$  
with $i \ne K$ and beyond is denoted as $V(a;\Omega)$.
Of course, in the Einstein-Hilbert limit the density~$\hat{\rho}_\mathrm{{ext}}$ is absent.

\medskip
Differentiating \eref{eq:Fr12a} with respect to the universal time $\tau$ yields, in analogy to~\eref{eq:f2a}, the acceleration equation
\begin{equation} \label{eq:ddot1}
	\ddot{a} = -\onehalf \frac{\d V(a;\Omega)}{\d a},
\end{equation}
\end{subequations}
which can be interpreted as the equation of motion (EOM) of a fictitious point particle of mass $m = 2$ in one-dimensional motion. 
Its ``velocity'' is $\dot{a}(\tau)$ where $\tau$ is the (dimensionless) time~\eqref{normaltime}. The term $V(a;\Omega)$ is interpreted as a given classical potential that has here been derived from the field equations of some semi-classical field theory in Lorentzian  spacetime.   
\eref{eq:Fr12a} (or~\eref{eq:f1a} for General Relativity) then appears as the conservation law for the "total energy", $\Omega_\mathrm{K}$,  consisting of the ``kinetic energy'' $\onehalf m \dot{a}^2$ and the ``potential energy'' $V(a;\Omega)$. 

\medskip
It is straightforward now to deduce along the lines of standard quantum mechanics of point particles that the Lagrangian, 
\begin{equation} \label{Lagrangian}
	\LCd(a,\dota;\Omega) :=  \onehalf m \dota^2 - V(a;\Omega),
\end{equation}
leads by variation of the corresponding action integral,
\begin{equation} \label{actionintegral}
	S_0(a) = \int \! \LCd(a,\dota;\Omega)\,\d \tau,
\end{equation}
to the EOM \eref{eq:ddot1}.
The associated Hamiltonian,  
\begin{subequations}
	\begin{alignat}{4}  
\HCd(p_a,a;\Omega) &= \dota \,p_a\,-\LCd, 
\label{HamiltonianDef}.
	\end{alignat}
becomes upon inserting the momentum dual to $\dota$  
	\begin{alignat}{4}  
		p_a :&= \pfrac{\LCd}{\dota } = m \dota = 2\dota \label{momentum}\\
  \HCd(p_a,a;\Omega) &=  \textstyle\frac{1}{2m} p_a^2 + V(a;\Omega)
		\label{Hamiltonian},
	\end{alignat}
\end{subequations}
and obviously satisfies the constraint equation
\begin{equation} \label{H=1/2K0}
	\HCd(p_a,a;\Omega) = \Omega_\mathrm{K}
\end{equation}
reflecting~\eref{eq:Fr12a}\footnote{Often the multiplication of~\eref{eq:HubbleGRextgen} by $a^p$ with $p > 2$ 
is applied that modifies not only the potential but also the canonical momentum and introduces ambiguities in the commutation relations. Then of course $m = m(a) =a^{p-2}$.}.

\subsection{The $\textbf{3}^{rd}$ quantization}

In the spirit of the Wheeler-DeWitt (WDW) approach \cite{wheeler57, DeWitt1967}, see also \cite{Kuchar:1991qf,Atkatz:1994hy, Kiefer:2008sw, He:2014yia, Giacosa:2018mkz},
we now consider the Universe to be a quantum mechanical object that, lacking any environment, will not by decoherence become a classical object. It is thus described by the wave function $\psi_\mathrm{U}(a,\tau;\Omega,\Omega_\mathrm{K})$ assumed to be a solution of a "quantum Friedman equation" that is obtained from of \eref{H=1/2K0} by canonical quantization: 
\begin{subequations} \label{def:canquant}
	\begin{alignat}{4} 
		p_a \rightarrow \hat{p}_a:&= -\im \frac{\d}{\d a} \label{momentumq}\\
		a \rightarrow \ahat :&=a \label{scaleq}\\
		\HCd(p_a,a;\Omega) \rightarrow \Hhat(\phat_a,\ahat;\Omega)&= \frac{1}{2m} \phat_a^2 + V(\ahat;\Omega). \label{Hamiltonianq}
	\end{alignat}
\end{subequations}
The quantized version of~\eref{H=1/2K0} thus leads to the Schr\"{o}dinger equation 
\begin{equation}
	\Hhat \, \psi_\mathrm{K}(a;\Omega) = \, \Omega_\mathrm{K}\,\psi_\mathrm{K}(a;\Omega). \label{Friedmanq}
\end{equation}
Of course, to specify a unique solution of this second-order PDE we need to fix the initial conditions of the wave function $\psi_\mathrm{K}$ and its first derivative, $\frac{\d \psi_\mathrm{K}}{\d a}$. This will be discussed in Sec.~\ref{sec:BC}.  

\medskip
The canonical quantization rules (\ref{def:canquant}) imply the following commutation relations:
\begin{subequations} \label{eq:commutators}
	\begin{alignat}{4} 
		\,[ \phat_a, \ahat ]\, &= -\im \label{com_pa}\\
		\,[ \phat_a^2, \ahat ]\, &= -2\im \phat_a \label{com_p2a}\\
		\,[ \phat_a^2, \phat\,\ahat ]\, &= 
		\phat_a \,[ \phat_a^2, \ahat ]\, =-2\im \phat_a^2 \label{com_p2pa}.
	\end{alignat}
\end{subequations}
According to the postulates of quantum mechanics, the wave function evolves in time by the unitary transformation 
\begin{equation}
    \Psi(a,\tau;\Omega) := \e^{-\im\,\Hhat\,\tau} \,\Psi(a,0;\Omega),
\end{equation}
which is equivalent to satisfying the time dependent version of the Schr\"{o}dinger equation \eqref{Friedmanq}, 
\begin{equation} 
	\Hhat \, \Psi(a,\tau;\Omega) = \, \im \pfrac{}{\tau}\,\Psi(a,\tau;\Omega), \label{Friedmanqt}
\end{equation}
with the eigenfunctions
\begin{equation} \label{def:eigenfunctions}
	\psi_\mathrm{K}(a,\tau;\Omega) = \psi_\mathrm{K} (a;\Omega) \, e^{-\im\tau\Omega_\mathrm{K} }.
\end{equation}
The Wheeler-DeWitt problem of missing time does not arise: Here the time variable $\tau$ is conjugate to the curvature parameter $K$ of the underlying FLRW geometry rather than to energy. 
If that curvature parameter, in this case the eigenvalue of the Hamiltonian $\Hhat$, is zero, then the corresponding eigenvector is time-independent, freezing the dynamics of the geometry\footnote{This is indeed what is often assumed in quantum cosmology studies including WDW.}.
Moreover, a spherical geometry ($K > 0$) propagating ``forward'' in time ("Universe"), i.e.\ with $\tau > 0$, is identical to a hyperbolic geometry ($K < 0$) moving ``backward'' in time ("Anti-Universe")\footnote{This remarkable asymmetry deserves further studies.}.

\medskip
The canonical quantization prescription thus defines a quantum theory if $\Hhat$ is self-adjoint with respect to the scalar product 
\begin{equation}
	\bra \psi | \phi \ket \,:=
	\int_{0}^{+\infty} \! \d a\,\,
	\psi^*(a)\,\phi(a),
\end{equation}
and its normalized and orthogonal eigenfunctions%
\footnote{Proof of \eref{eq:ONS}:
	\begin{align*}
		&\Omega_\mathrm{K}{^\prime}\int \! \d a\, 
		\psi_\mathrm{K}^*(a;\Omega)\,\psi_{K^\prime}(a;\Omega) = 
		\int \! \d a\, 
		\psi_\mathrm{K}^*(a;\Omega)\,\Hhat\,\psi_{K^\prime}(a;\Omega) \\
		&= \int \! \d a\, 
		\left(\Hhat\,\psi_\mathrm{K}(a;\Omega)\right)^*\,\psi_{K^\prime}(a;\Omega) 
		= \Omega_\mathrm{K} \int \! \d a\, 
		\psi_\mathrm{K}^*(a;\Omega)\,\psi_{K^\prime}(a;\Omega) \nonumber 
	\end{align*}
	and thus 
	\begin{align*}
		\left( \Omega_\mathrm{K}{^\prime} - \Omega_\mathrm{K}\right) 
		\int \! \d a\, 
		\psi_\mathrm{K}^*(a;\Omega)\,\psi_{K^\prime}(a;\Omega) = 0 
	\end{align*}
	which for $\Omega_\mathrm{K}{^\prime} \ne \Omega_\mathrm{K}$ implies the orthogonality relation.}
 facilitate a complete basis of the Hilbert space\footnote{We skip the proof of completeness here and address the property of the momentum operator on the half-line in Section \ref{sec:BC}.}: 
\begin{subequations}
\begin{alignat}{4}
\label{eq:ONS}
	\bra K | K^\prime \ket &= \delta_{KK^{\prime}} \\
	\label{def:complete}
	\Hhat &\equiv\int\kern-1.5em\sum_{\,\,\,K}  \Omega_\mathrm{K}\,|K \ket \bra K|.
\end{alignat}
\end{subequations} 
Notice that $K$ stands here for $\Omega_\mathrm{K}$, with $\int_K \sim \int \!\d \Omega_\mathrm{K}$ or $\sim \sum_K \Omega_\mathrm{K}$ for continuous or discrete portions of the spectrum.
For any two time-dependent solutions, $\psi$ and $\phi$, of \eref{Friedmanqt} it is also straightforward to prove  the continuity equation
\begin{equation} \label{eq:continuity}
	\pfrac{}{\tau}\,P(a,\tau) + \pfrac{}{a} J(a, \tau) = 0
\end{equation}
with  
\begin{subequations}
	\begin{alignat}{4}  
		P(a,\tau) &:= \psi^*(a,\tau)\,\phi(a,\tau) \\
		J(a,\tau) &:= - \textstyle\frac{i}{2m} \,\left[
		\psi^*(a,\tau)\,\pfrac{}{a}\,\phi(a,\tau)
		- \left(\pfrac{}{a}\,\psi^*(a,\tau)\right)\,\phi(a,\tau)
		\right].
	\end{alignat}
\end{subequations}

Let $\psi_\mathrm{U}$ denote the normalized universal wave function, the probability amplitude of the Universe's geometry in the parameter space $\left\{a,\Omega_\mathrm{K} \right\}$ of the mini-superspace \cite{Atkatz:1994hy}. The normalization condition,
\begin{equation} \label{def:normalization}
	\int \! \d a\, 
	\psi_\mathrm{U}^*(a,\tau;\Omega)\,\psi_\mathrm{U}(a,\tau;\Omega) =: \bra U\,|\,U \ket |_\tau = 1,
\end{equation}
expresses the assumption that the Universe exists at any time\footnote{At least as long as time remains a meaningful concept.}.

\medskip
In the Copenhagen interpretation of quantum mechanics, a observable is expressed by a self-adjoint operator, say $\hat{\OCd}$.
The likelihood for a value of a measurement of that observable is then defined by the expectation value  
\begin{equation}
	\braU\hat{\OCd}\Ukett := \int \! \! \d a \,\psi_\mathrm{U}^*\,\hat{\OCd} \,\psi_\mathrm{U} =: \int \! \! \d a \,\OCd_U(a,\tau).
\end{equation}
(Notice that for the sake of brevity we suppress below the involved parameters $\Omega$ and $K$.) 
As is found in any textbook on quantum mechanics, the dynamics and the uncertainty relation of arbitrary quantum observables, say $\hat{\OCd}, \hat{\PCd}$ and $\hat{\QCd}$, can readily be derived from the canonical commutation relations:
\begin{subequations}
	\begin{alignat}{4} 
		\frac{\d}{\dtau}\!\bra\hat{\OCd}\ket |_\tau &= \im\bra\,[\Hhat,\hat{\OCd} ]\,\ket |_\tau + \bra \pfrac{\hat{\OCd}}{\tau}\ket |_\tau \label{eq:dotO}\\
		\left(\Delta \OCd \right)^2 &:= \, \bra\left(\hat{\OCd}-\bra\hat{\OCd}\ket |_\tau\right)^2\ket \,\equiv \,\bra \hat{\OCd}^2\ket |_\tau - \bra\hat{\OCd}\ket^2 |_\tau \label{def:uncertainty}\\
		\left(\Delta \PCd \right) \left(\Delta \QCd \right) &\ge \onehalf \left|\bra\,[\hat{\PCd},\hat{\QCd}]\,\ket |_\tau\right|.  \label{eq:uncertrel}
	\end{alignat}
\end{subequations}
\eref{eq:uncertrel} immediately gives with \eref{com_pa} Heisenberg's uncertainty relation,
\begin{equation}
	\left(\Delta p_a \right) \left(\Delta a \right) \ge \onehalf.  \end{equation}

\medskip
The wave function of the Universe, $\psi_\mathrm{U}(a,\tau ; \Omega)$, can be expanded in terms of the eigen-basis~\eqref{def:eigenfunctions}  of the Hamiltonian~\eqref{Hamiltonianq}:
\begin{equation} \label{eq:expansion}
	\psi_\mathrm{U}(a,\tau;\Omega) = \int\kern-1.5em\sum_{\,\,\,K} c_\mathrm{U}(\Omega_\mathrm{K})\,\psi_\mathrm{K}(a;\Omega)\, \, e^{-\im\,\Omega_\mathrm{K}\,\tau}.
\end{equation}
The expansion coefficients are, by completeness and ortho-normality of the basis, calculated as
\begin{equation}
    c_\mathrm{U}(\Omega_\mathrm{K}) := \int \! \! \d a \, \psi^*_\mathrm{K}(a;\Omega) \,\psi_\mathrm{U}(a,\tau=0;\Omega_\mathrm{g}),
\end{equation}
and, combined with \eref{def:normalization}, satisfy the normalization condition
\begin{equation} \label{def:norm}
	\int\kern-1.5em\sum_{\,\,\,K} c^*_\mathrm{U}(\Omega_\mathrm{K})\, c_\mathrm{U}(\Omega_\mathrm{K}) = 1.
\end{equation}
$|c_\mathrm{U}(\Omega_\mathrm{K})|^2$ is then the q.m. probability for the Universe to be in the state $\psi_\mathrm{K}$. 
 The quantum-mechanical expectation value of the scale parameter $\am = \bra \ahat \ket$ is thus the function 
\begin{align} \label{expvalue}
	\am(\tau) &:= \bra U|\ahat|U\ket |_\tau = \int \!da \, 
	\psi_\mathrm{U}^*(a,\tau;\Omega)\,a\, \psi_\mathrm{U}(a,\tau;\Omega) \\
	&= \,\int\kern-1.5em\sum_{\,\,\,K} \,\int\kern-1.5em\sum_{\,\,\,K^{\prime}}\,c^*_\mathrm{U}(\Omega_\mathrm{K}) \,c_\mathrm{U}(\Omega_\mathrm{K}') 
	\,e^{-\im(\Omega_\mathrm{K}^\prime - \Omega_\mathrm{K}) \tau}
	\,\int \!da\,\psi_\mathrm{K}^*(a;\Omega)\,a\,\psi_{K^\prime}(a;\Omega) \nonumber\\
	&=\, \int\kern-1.5em\sum_{\,\,\,K} \,\int\kern-1.5em\sum_{\,\,\,K^{\prime}}\,\rho_{KK^{\prime}}\,\bra K|a|K^\prime\ket\,e^{-\im(\Omega_\mathrm{K}^\prime - \Omega_\mathrm{K})\tau},
	\nonumber
\end{align}
with the density matrix $\rho_{KK^{\prime}} := c^{\textcolor{red}{*}}_\mathrm{U}(\Omega_\mathrm{K}) \,c_\mathrm{U}(\Omega_\mathrm{K}')$ and the scale transition matrix $\bra K|a|K^\prime\ket := \int \!da\,\psi_\mathrm{K}^*(a;\Omega)\,a\,\psi_{K^\prime}(a;\Omega)$. 
Notice that if the wave function of the Universe were an eigenfunction of the Hamiltonian, $\am(\tau) = const$ would follow\footnote{This is readily seen if that eigenvalue is $\bar{K}$ by setting $c^*_\mathrm{U}(\Omega_\mathrm{K}) = \delta_{K\bar{K}}$ in \eref{expvalue}.}, and the Universe were stationary. The value of the present time $\tau_0$ is extracted from $\am(\tau_0) = a_0$\footnote{We reinstate $a_0$ here for later use. Notice that the classical calculations listed in Section \ref{sec:CCGG} remain unchanged if $a/a_0$ is substituted for $a$ there.}. 

\medskip
The expectation value of the metric curvature, $\Omega_\mathrm{K}$, is equivalent to the expectation value of the "time operator" $\partial / \partial \tau$, 
\begin{align}
	\bra U|\,\im\pfrac{}{\tau}\,|U\ket |_\tau &= \bra U|\,\Hhat \,|U\ket |_\tau \\
	&= \int \! \d a \,\psi_\mathrm{U}^*
	\Hhat
	\psi_\mathrm{U}(a,\tau;\Omega) \nonumber \\
	&= \, \int\kern-1.5em\sum_{\,\,\,K} \,\int\kern-1.5em\sum_{\,\,\,K^{\prime}}\,\rho_{KK^{\prime}}\,\bra K|\Omega_{K^\prime}|K^\prime\ket\,e^{-\im(\Omega_\mathrm{K}^\prime - \Omega_\mathrm{K})\tau} \nonumber \\
	&=
	\int\kern-1.5em\sum_{\,\,\,K} \Omega_\mathrm{K}\,c^*_\mathrm{U}(\Omega_\mathrm{K}) \, c_\mathrm{U}(\Omega_\mathrm{K}) 
	=: \, 
	\Omega_{\bar{\mathrm{K}}}
	= \text{const.},
	\nonumber
\end{align}
and is thus stationary. However, if the observed flatness of the current Universe is to be reproduced, then $\Omega_{\bar{K}} \approx 0$. Furthermore, the normalized Hubble function  $E^2(\tau) := \left(H(t)/H_{100}\right)^2$, see~\eref{def:E(tau)}, should be recovered as the expectation value of 
\begin{equation} \label{def:E2}
	E^2(\tau) 
	=  -\braU\frac{V(\ahat;\Omega_i) - \im\pfrac{}{\tau}}{\ahat^2}\Uket |_\tau
	=  -\quarter \braU \frac{\phat^2}{\ahat^2} \Ukett
	,
\end{equation}
that at the present time $\tau_0$ should satisfy 
\begin{equation} \label{expvalue2}
	E^2(\tau_0) = h^2.
\end{equation}
Proving in this way the model's consistency will require both, a proper definition of the ordering\footnote{Depending on some selection of the operator ordering, additional, more or less arbitrary ``quantum corrections'' might thus arise in standard quantum mechanics.} of the operator $\phat^2/\ahat^2$, and tuning the  parameters $\Omega$. 
\subsection{Initial conditions for the universal wave function}\label{sec:BC}

The discussion of proper initial conditions for the universal wave function is still ongoing~\cite{Atkatz:1994hy}. 
Hartle and Hawking proposed the so called no-boundary wave function \cite{Hartle:1983ai} which is a real function facilitating both, outgoing and ingoing waves\footnote{From \eref{eq:schroedinger} the quantum potential must then, with $S \equiv 0$, just cancel the classical one.}. 
The Vilenkin version \cite{vilenkin15}, in contrast, considers the birth of the Universe by quantum tunneling. 
The solution is an outgoing wave into the classically allowed region outside the Planck-size environment of the singularity at the origin. 
Both approaches rely on the first order WKB approximation of the timeless Wheeler-De Witt equation, assuming the Universe to rapidly become classical. 
The applicability of this approximation scheme is not {a} priori clear. Berkowitz \cite{Berkowitz:2020asv}, to quote another example, sets for the amplitude $R(0) =1$ and $R^\prime(0) = -1$. 

\medskip
We wish, for reasons addressed above, to suppress the Big Bang singularity at the origin by requesting that the probability density vanishes there \cite{Kasem:2020wsp}.  
Thus we would like to stipulate for any eigensolution $\psi_K(a;\Omega) = R \, \exp{(iS)}$ DeWitt's criterion \cite{DeWitt1967}:

\begin{equation}
	R(a=0)
	= 0.  \label{BC1}
\end{equation}
The symmetry of the momentum operator can be ensured with
\begin{equation}
	R(a=0) = R(a=\infty)	= 0.  \label{BC2}
\end{equation}
However, it is well known that the momentum operator $p_a$ with this  boundary condition is not self-adjoint on the half-line $[0,\infty)$. Thus in order to rigorously ensure the self-adjointness of $\hat{p}_a$, we impose the cyclic boundary condition, 
\begin{equation}
	R(a=0) 
	= R(a=L) = 0,  \label{BC3}
\end{equation}
with some length parameter (cutoff) $L$ that can subsequently be sent\footnote{Numerically this concept does not appear efficient, though.} to $\infty$.

\medskip
An approach along the lines of the "no-boundary" approach \cite{Hartle:1983ai} is  particularly suitable for numerical analyzes. 
The asymptotic behavior of the wave function can be determined from the asymptotic form of the potential. The solution basis found in this way can subsequently be combined to a new basis satisfying the desired initial conditions. 
That numerical study together with an analysis of the boundary conditions will be discussed in a follow-up paper. 

\subsection{The measurement of the Universe} \label{sec:EI}
In  quantum cosmology we are facing the problem that the standard, Copenhagen notion of quantum mechanical measurement, postulating a collapse of the wave function to an eigenstate of the operator representing the measuring device, is not applicable. 
Adhering to that interpretation of quantum mechanics  requires separating the Universe into an observed system, the Universe itself, and a macroscopic observer. 
In this situation, though, the observer is a part of the observed object, and it is very unlikely that any measurement would "collapse" the Universe into a new state. 

\medskip
To formulate what the wave function and a measurement of the properties of the Universe could be we follow the line of thought of Refs. \cite{durr1992quantum2, durr2004quantum}.  
The observed portion of the Universe is then described by a universal wave function that is a superposition of all space-time geometries within the selected mini-superspace 
characterized by a curvature eigenvalue $\Omega_\mathrm{K}$ and the scale factor $a$:  
\begin{equation} \label{eq:expansion2}
 \psi_\mathrm{U}(a,\tau;\Omega) = \int\kern-1.5em\sum_{\,\,\,K} c_\mathrm{U}(\Omega_\mathrm{K})\,\psi_\mathrm{K}(a;\Omega)\, \, e^{-\im\,\Omega_\mathrm{K}\,\tau}.
\end{equation}
The total wave function must include also the "environment" pertinent to each observer, i.e. the wave function of the measurement device, the \emph{apparatus} composed of observatories, computers, software, people etc. with a microscopic, potentially very complex configuration described by the coordinate set $\{y\}$. (For simplicity we will in the following use the notation  $y$ instead of $\{y\}$.) This apparatus is calibrated such that for a particular  configuration, say $y_K$, it delivers a particular measurement result, the "pointer reading", $\Omega_\mathrm{K}$, of the spatial curvature. Be this abstract "calibration" of the apparatus expressed by the relation $F(y)$ with $F(y_K) = \Omega_\mathrm{K}$, and $F^{-1}(\Omega_\mathrm{K}) = y_K$. For the apparatus to deliver unambiguous results, i.e. with different configurations giving rise to unique pointer readings, we must require that for $\Omega_\mathrm{K} \ne \Omega_{K'}$ the respective supports $F^{-1}(\Omega_\mathrm{K}) = y_K$ and $F^{-1}(\Omega_{K'}) = y_{K'}$ are disjoint. (Of course, for a continuous spectrum $\{\Omega_\mathrm{K}\}$ this can be valid only approximately, but we can assume the supports to be sufficiently disjoint to align with the given accuracy of the apparatus.) 

\medskip
Assuming -- rather reasonably -- that the Universe remains unaffected by all observations conducted by any such apparatus, the pertinent measurement processes  must be  "weak" \cite{durr2004quantum} or "subquantum" \cite{hiley2006delayed}. 
Be 
\begin{equation}
\mathbf{\Psi}_\mathrm{tot}(a,\tau_b, y) = \psi_\mathrm{U}(a,\tau_b) \, \Phi_0(a,y,\tau_b). 
\end{equation}
the total wave function at time $\tau_b$, before any of the measurements have taken place. 
$\Phi_0(a,y,\tau_b)$ is the pre-measurement state and the starting point for the series of configurations $y_n$ of any apparatus used across the history of $N$ observations, $n = 1, 2, 3, ...N < \infty$.
$\psi_\mathrm{U}$ is given by \eref{eq:expansion2}.  After completing a series of weak measurement events at time $\tau \gg \tau_b$ the total wave function is transformed to a new state by the unitary transformation
\begin{align}
\mathbf{\Psi}_\mathrm{tot}(a,\tau, y) &= U(\tau - \tau_b)\,\mathbf{\Psi}_\mathrm{tot}(a,\tau_b, y) \\
&=
\int\kern-1.5em\sum_{\,\,\,K} c_\mathrm{U}(\Omega_\mathrm{K})\,\psi_\mathrm{K}(a)\, \, e^{-\im\,\Omega_\mathrm{K}\,\tau} \,\sum_n \Phi_\mathrm{K}(a,y_n,\tau) .\nonumber
\end{align}
While the "background" Universe remains unaffected by the  independent (parallel or subsequent) measurements\footnote{Here we implicitly assume the system to be ergodic as averaging in time over a series of measurements performed at different albeit adjacent instants is identified with averaging over an ensemble of identical subsystem \cite{Shtanov:1995ie}.}, the "observers" wave packet spreads across many possible states representing all possible pointer readings denoted by (the index) $K$. By the above construction each  apparatus will be assigned the states $\Phi_\mathrm{K}(y_n,\tau)$ representing the readings $\Omega_\mathrm{K_n}$ only on the "sufficiently" disjoint supports\footnote{A possible representation of such a wave packet could be for example (for a given $a$ and with the time dependence suppressed)
\begin{equation*}
    \Phi_\mathrm{K}(y_n) = C_n\,\exp{\left[- \left( \frac{y-y_n}{2\sigma_n} \right)^2 \right]}\delta_{K K_n}
\end{equation*}
where the width $\sigma$ of the Gaussian can be made as narrow as  the accuracy of the individual experimental setup requires. In that limit the series of experiments is simplified  to deliver distinct results with vanishing error bars.} $F^{-1}(\Omega_\mathrm{K_n})$ for different $\Omega_\mathrm{K_n}$  \cite{durr1992quantum2}. For $n \ne m$ and $K \ne K'$ we thus expect
\begin{equation} \label{def:support}
    \Phi_\mathrm{K}(a,y_n,\tau)\,\Phi_\mathrm{K'}(a,y_m,\tau) \approx 0.
\end{equation}
The quantum mechanical probability for the pointer reading $\Omega_{{K_n}}$, 
\begin{equation}
p_{{n}} = \int \!\!\! \d a \int_{F^{-1}({K_n})} \,\!\! \d y \,
\left|\mathbf{\Psi}_\mathrm{tot}(a,\tau, y)\right|^2,
\end{equation}
simplifies now considerably when taking into account the property (\ref{def:support}) of the apparatus wave function, giving, in the idealized case of "sharp" measurements,
\begin{align*}
&\int_{F^{-1}({K_n})} \!\!\! \d y \,
\left|  \Phi^*_\mathrm{K'}(a,y,\tau) \, \Phi_\mathrm{K''}(a,y,\tau)  \right| = \\ 
& \int \!\!\! \d y_n \,
\left|  \Phi^*_\mathrm{K'}(a,y_n,\tau) \, \Phi_\mathrm{K''}(a,y_n,\tau)  \right|
\approx    \delta_{K_n K'}\,\delta_{K_n K''}.\nonumber
\end{align*}
With the normalization condition \eqref{eq:ONS} this finally gives
\begin{equation}
    p_{{n}} = \left|c_\mathrm{U}(\Omega_\mathrm{K_n})\right|^2,
\end{equation}
This result is indistinguishable from the "collapse" paradigm of the orthodox Copenhagen interpretation \cite{von2013mathematische}.

\medskip
 The "weak" measurements can now be used to restrict the Hilbert space to an empirically determined subspace in which an \emph{effective wave function} is located. 
 Consider the system wave function at time $\tau$, after all $N$ measurements  have been performed:   
\begin{equation} \label{def:effectiveWF}
    \psi_\mathrm{U}(a,\tau) \rightarrow
    \sum_n \,c_\mathrm{U}(\Omega_\mathrm{K_n})\,\psi_\mathrm{K_n}(a)\, \, e^{-\im\,\Omega_\mathrm{K_n}\,\tau}.
\end{equation}
 This reflects our expectation -- the more we learn through observations, the more accurate our description of the Universe, 
 i.e.\ its wave function, becomes.

\medskip
Since each observer, i.e.\ any ``measurement apparatus'', is an imperfect device, the measurement values will not be a discrete set. Each measurement will rather be distributed around the mean $\bar{K}_n$ within some more or less narrow interval $\Omega_\mathrm{K} - \Omega_{\bar{K}_n} \lesssim \Delta_n$ (``error bars''). This relieves the idealized assumption  of "sharp" results per experiment used above for simplicity. The memory of all past measurements can then be imprinted on the expansion coefficients via 
\begin{equation}
 c_\mathrm{U}(\Omega_\mathrm{K_n}) \rightarrow c_{\mathrm{U},n}({K}) = q_\text{n}(K)\, \NCd(\Omega_\mathrm{K};\Omega_\mathrm{\bar{K}_\text{n}},\sigma^2_n),
 \end{equation}
 where 
 \begin{equation}
     \NCd(\Omega_\mathrm{K};\Omega_\mathrm{\bar{K}_\text{n}},\sigma^2_\text{n})
= \frac{1}{\sqrt{2\pi \sigma_\text{n}^2}}\,
 \exp{\left[-\onehalf \left(\frac{\Omega_\mathrm{K}-\Omega_\mathrm{\bar{K}_\text{n}}}
 { \sigma_\mathrm{\bar{K}_\text{n}} } \right)^2
 \right]
 }
\end{equation}
is the Gaussian normal distribution.
The expansion coefficients $q_\text{n}(K)$ quantize the relative contribution of the $\text{n}^{th}$ measurement of $\Omega_\mathrm{K}$, normalized in line with \eref{def:norm}\footnote{This form of the wave function would also be used in Everett's many-worlds theory  \cite{everett1957relative, Tegmark:2009pj} but the interpretation of the measurement process is entirely different \cite{valentini2010broglie}.}. 
This now fixes the effective wave-function of the space-time geometry in a homogeneous and isotropic Universe (with matter in form of ideal fluids) to the empirical, effective form:
\begin{equation} \label{eq:expansion3}
 \psi_\mathrm{U}(a,\tau;\Omega) = 
 \int\kern-1.5em\sum_{\,\,\,K} \,\sum_n^N
 q_\text{n}(K)\,
 \NCd(\Omega_\mathrm{K};\Omega_\mathrm{\bar{K}_\text{n}},\sigma^2_\text{n})
 \,\psi_\mathrm{K}(a;\Omega)\, \, e^{-\im\,\Omega_\mathrm{K}\,\tau}.
\end{equation}
By the fact that a finite superposition of Gaussians is again a Gaussian, 
\begin{subequations} 
\begin{align}
   \sum_n^N \,q_\text{n}\,\NCd(\Omega_\mathrm{K};\Omega_\mathrm{\bar{K}_n},\sigma^2_\text{n}) 
&=
\NCd(\Omega_\mathrm{K};\Omega_\mathrm{\bar{K}},\sigma^2),
\end{align}
where
\begin{align}
   \Omega_\mathrm{\bar{K}} &= \sum_n^N q_\text{n}\,\Omega_\mathrm{\bar{K}_n}\\\sigma^2 &= \sum_n^N q^2_\text{n}\,\sigma^2_\text{n},
\end{align}
\end{subequations}
\eref{eq:expansion3} can be consolidated to the effective (normalized) wave function
\begin{equation} \label{eq:expansion31}
 \psi_\mathrm{U}(a,\tau;\bar{K},\Omega) = 
 q_{\bar{\mathrm{K}}}\,\int\kern-1.5em\sum_{\,\,\,K} \,
  \e^{-\left( \frac{\Omega_\mathrm{K}-\Omega_{\bar{\mathrm{K}}}}{ 2\sigma} \right)^2}
 \,\psi_\mathrm{K}(a;\Omega)\, \, e^{-\im\,\Omega_\mathrm{K}\,\tau}.
\end{equation}
We further observe, via \eref{eq:HubbleGRextgen}, that (up to the sign) the parameter $h$ is equivalent 
to $\Omega_K$:
\begin{equation} \label{def:Omega0}
    \Omega_\mathrm{K}(h) = h^2 -  \Omega_\mathrm{m} - \Omega_\mathrm{r} - \Omega_\Lambda - \hat{\rho}_\mathrm{ext}(a=1) . 
\end{equation}
Substituting then $\Omega_K = \Omega_K(h)$, i.e. $K = K(h)$, into \eref{eq:expansion31}, the effective wave function can be re-written as  
\begin{equation} \label{eq:expansion4}
\Psi_\mathrm{U}(a,\tau;\bar{h},\Omega) = 
 \frac{1}{C}\,\int \!\d h^2 
 \,\e^{-\left(
 \frac{
 h^2-\bar{h}^2
 }
 { 2\sigma }
 \right)^2
 }
 \,\psi_{h}(a;\Omega)\, \, e^{-\im\,(h^2-\bar{h}^2)\,\tau}.
\end{equation}
where the quantity $\bar{h}$ is the relative Hubble constant for the FLRW spatial curvature $\bar{K}$.
Here $\psi_{h}(a;g_1)$ are still eigensolutions of the eigenvalue problem \eqref{Friedmanq} that by a similar substitution of parameters is recast to
\begin{equation}
	\Hhat \, \psi_h(a;\Omega) 
	= \, \left(h^2-\bar{h}^2 \right)\,\psi_h(a;\Omega). \label{Friedmanq2}
\end{equation}

\subsection{The polar representation } \label{sec:eikonal}
Let the universal wave function be expressed in the polar representation as
\begin{equation}
	\psi_\mathrm{U}(a,\tau) =  R_\mathrm{U}(a,\tau)\,e^{\im S_\mathrm{U}(a,\tau)},
\end{equation}
where $R_\mathrm{U}$ and $S_\mathrm{U}$ are w.l.o.g. real functions representing the amplitude and modulus (phase), respectively. Notice that $S_\mathrm{U} = S_\mathrm{U}$~mod~$2\pi$ and $m=2$.
Inserted into the Schr\"{o}dinger equation~\eqref{Friedmanqt} we get for the real and imaginary part, and $R_\mathrm{U} \ne 0$, respectively,
\begin{subequations}
	\begin{alignat}{4} 
		&\pfrac{S_\mathrm{U}}{\tau} + \frac{1}{2m} \left(\pfrac{S_\mathrm{U}}{a}\right)^2 + V(a) 
		- \frac{1}{2mR_\mathrm{U}}
		\frac{\partial^2 R_\mathrm{U}}{\partial a^2}  = 0 \label{eq:schroedinger}\\
		&\pfrac{R^2_\mathrm{U}}{\tau} + \frac{1}{m}\pfrac{}{a} \left(R_\mathrm{U}^2 \pfrac{S_\mathrm{U}}{a} \right) = 0. \label{eq:probcons}
	\end{alignat}
\end{subequations}
The last term in the Schr\"{o}dinger equation~(\ref{eq:schroedinger}),
\begin{equation} \label{def:Vquant}
 V_{quant} :=  - \frac{1}{2m}	\frac{R_\mathrm{U}''}{R_\mathrm{U}},
\end{equation}
(the prime denotes derivation w.r.t. $a$) is the \emph{quantum potential}, correcting the classical potential $V(a)$. 
\eref{eq:probcons}, in addition,  
reproduces the continuity equation \eqref{eq:continuity} with $\psi = \phi = \psi_\mathrm{U}$ and
\begin{subequations}
	\begin{alignat}{4} 
		P_\mathrm{U} &= R_\mathrm{U}^2 =: \rho_\mathrm{U} \label{probabilitydensity} \\
		J_\mathrm{U} &= \frac{1}{m} R_\mathrm{U}^2\,\pfrac{}{a} S_\mathrm{U} =: 
        \rho_\mathrm{U}\,\nu_\mathrm{U}. \label{current1}
	\end{alignat}
\end{subequations}
$\rho_\mathrm{U}(a,\tau)$ represents the quantum mechanical probability density for measuring the position of the fictitious particle (scale of the mini-superspace geometry), $a$, at ``time''~$\tau$, and 
\begin{equation}
    \nu_\mathrm{U}(a,\tau) := \frac{1}{m}\,\pfrac{}{a} S_\mathrm{U}(a,\tau)
\end{equation}
is the pertinent velocity field.
Since $J_\mathrm{U}(a,\tau)$ is real and $\phat_a$ is assumed to be a self-adjoint operator,  
\begin{equation}
	J_\mathrm{U}(a,\tau) = \frac{1}{m} \,\psi_\mathrm{U}^*\,\phat_a \,\psi_\mathrm{U} =: \frac{1}{m} \pi_\mathrm{U}(a,\tau).
\end{equation}
The continuity equation, now expressed as 
\begin{equation} \label{consprobab}
    \pfrac{}{\tau}\,\rho_\mathrm{U}(a,\tau)\,
    + m \,\pfrac{}{a} \left(\rho_\mathrm{U}(a,\tau)\,\nu_\mathrm{U}(a,\tau)\right) =0,
\end{equation}
can be interpreted as the \emph{conservation law of probability}, where the probability density current  is 
\begin{equation} \label{eq:equilhypothesis}
	\pi_\mathrm{U} \equiv\, \rho_\mathrm{U}\,\pfrac{S_\mathrm{U}}{a} 
    = m \rho_\mathrm{U}\, \nu_\mathrm{U}. 
\end{equation}
In terms of the wave function the modulus ("eikonal" in wave optics) and the q.m. probability velocity are, respectively: 
\begin{subequations}
\begin{alignat}{4}
    \rho_\mathrm{U}(a,\tau) &= \psi^*_U \,\psi_\mathrm{U} = R_\mathrm{U}^2 \label{def:prob}\\
    S_\mathrm{U}(a,\tau) &= -\ihalf \,\ln \left(\psi_\mathrm{U} / \psi^*_U \right) \label{def:eikonal}\\
    \nu_\mathrm{U}(a,\tau) &= \frac{1}{m} \, \IM \left( \frac{\partial \psi_\mathrm{U} / \partial a}{\psi_\mathrm{U}}\right) 
    \label{def:probvelo}
        \end{alignat}
\end{subequations}

\medskip
From Eqs.~\eqref{eq:dotO} and \eqref{current1} we now find for the quantum mechanical expectation value of the expansion rate
\begin{align}
	\dot{\am}(\tau) :&= \frac{\d}{\dtau} \braU\ahat\Uket |_\tau 
    = \im \,\braU\,[\Hhat,\ahat]\,\Uket |_\tau \nonumber \\ 
	&= \im \,\braU\,[\frac{1}{2m}  \, \phat_a^2,\ahat]\,\Uket |_\tau \\
	&= \frac{1}{m} \,\braU\, \phat_a\,\Uket |_\tau 
	= \frac{1}{m}  \,\int \! \! \d a\,\rho_\mathrm{U}(a,\tau) \pfrac{S_\mathrm{U}(a,\tau)}{a}.
	\nonumber
\end{align}
The acceleration, the rate of change of the expansion rate, is then given by
\begin{align}
	\ddot{\am}(\tau) :&= \frac{1}{m}  \frac{\d}{\dtau} \braU\phat_a\Ukett 
	= \frac{\im}{m}  \,\braU\,[\Hhat,\phat_a]\,\Uket |_\tau \\
	&= \frac{\im}{m}  \,\braU\,[V(\ahat),\phat_a]\,\Uket |_\tau 
	= -\onehalf \int \! \! \d a\,\rho_\mathrm{U}(a,\tau) \frac{\d V(a)}{\d a},
	\nonumber
\end{align}
in analogy to the Friedman equation \eqref{eq:ddot1} (``Ehrenfest theorem''). (Recall that the mass $m=2$ of the fictitious particle in this one-dimensional dynamics gives just the factor $\nicefrac{1}{2}$  between momentum and velocity.)

\subsection{The classical limit}

When the quantum correction potential $V_{quant}$ in the formulation~(\ref{eq:schroedinger}) of the Schr\"{o}dinger equation can be neglected, in other words, if the condition
\begin{equation} \label{cond:classicality}
     |R_\mathrm{U}''| \ll |R_\mathrm{U}| 
\end{equation}
applies, we get 
\begin{equation}
    \pfrac{S_\mathrm{U}}{\tau} + \frac{1}{2m} \left(\pfrac{S_\mathrm{U}}{a}\right)^2 + V(a) 
		 = 0. \label{eq:HamJac}
\end{equation}
This equation is formally identical to the Hamilton-Jacobi equation, 
\begin{equation}
       \pfrac{S}{\tau} + H_\mathrm{Cl}(\pfrac{S}{a},a) = 0, \label{eq:HamJac2} 
\end{equation}
of classical point mechanics \cite{greiner10}. The phase $S_\mathrm{U}(a,\tau,\Omega_\mathrm{K})$ is now identified with the  generating function  $S(a, P_a,\tau)$ of a canonical transformation (of the second kind) that transforms the classical momentum $p_a = \dot{a}$ into a new momentum $P_a$, and the classical Hamiltonian $H_\mathrm{Cl}(p_a,a)$ into a new Hamiltonian that satisfies the boundary condition $H'(P_A,A) \equiv 0$. Via the canonical equations
\begin{align*}
\pfrac{H'(P_A,A)}{A} &= -\dot{P_A} \equiv 0 \\
\pfrac{H'(P_A,A)}{P_A} &= \dot{A} \equiv 0,
\end{align*}
we find the new phase-space variables to be constant, i.e. we can set $P_A \equiv \alpha$ and $A \equiv \beta$. These constants are yet undefined and will be fixed by the initial condition. The generating function of the second kind links the old and new phase-space variables to the original ones as  
\begin{align*}
\pfrac{S}{a}&= p_a \\
\pfrac{S}{P_a}&= A = \beta.
\end{align*}
Since the explicit $\tau$ dependence is present only in the first term of \eref{eq:schroedinger}, we can make the linear ansatz 
\begin{equation*}
    S(a,\alpha,\tau) = W(a,\alpha) - \alpha \tau.
\end{equation*}
Then \eref{eq:schroedinger} becomes (with $m=2$ reinstated)
\begin{equation}
    -\alpha + \quarter \left(\pfrac{W}{a}\right)^2 + V(a) = 0,
\end{equation}
which gives
\begin{equation}
    W = 2\int^a_{a_0} \d a' \sqrt{\alpha- V(a')}
\end{equation}
and thus
\begin{equation}
    \beta = \pfrac{S}{\alpha} = \int^a_{a_i} \frac{\d a'}{\sqrt{\alpha- V(a')}} - \tau. 
\end{equation}
If we apply now the boundary condition $a = a_0$ for the scale parameter at the present time $\tau_0$, and set the initial scale to $a = a_i$ at some instant $\tau_i$,  we get
\begin{equation}
    \tau_0 + \beta = 
    \int^{a_0}_{a_i} \frac{\d a'}{\sqrt{\alpha- V(a')}}.
\end{equation}
Setting in addition the "age of the Universe" (aka event horizon) equal to the  Hubble time,  
\begin{equation} \label{def:Hubbletime}
t_0 = 1/H_0, \qquad \tau_0 = t_0\,H_{100}=\frac{1}{h},    
\end{equation}
gives $\beta \equiv -\tau_i$. 
(Notice that the result is independent of the value of the present scale parameter $a_0$ which can for simplicity be set to $a_0 = 1$.) 

\medskip
We note in passing that the physical significance of the generating function $S$ derives from the following relation for the generating function,
\begin{equation*}
    \frac{dS(a,P_A,\tau)}{d\tau} = \pfrac{S}{\tau} + \pfrac{S}{a}\,\dot{a} = p_a\,\dot{a} - H_\mathrm{Cl} = L_\mathrm{Cl}, 
\end{equation*}
from which immediately follows
\begin{equation*}
    S = \int^\tau \dtau' L_{Cl}(a,\dot{a},\tau') \equiv S_\mathrm{Cl}(\tau).
\end{equation*}
$S$ obviously corresponds to the action integral $S_\mathrm{Cl}$, recovering the definition of the elementary probability amplitude in quantum mechanics.
The Schr\"{o}dinger equation \eqref{eq:schroedinger} of quantum cosmology becomes, under the classicality condition~\eqref{cond:classicality}, the Hamilton-Jacobi problem of classical point dynamics\footnote{This relation, and the relation to ray optics, was indeed the starting point also for Schr\"{o}dinger when developing his famous equation.}.

\section{The de Broglie-Bohm approach} \label{sec:debrogliebohm}

In this Section we complement the discussion in~\sref{sec:EI} of the weak measurement framework by an alternative interpretation of quantum mechanics, the pilot-wave formalism developed by Louis de Broglie and David Bohm~\cite{Bohm:1951xx, Bohm:1951xw, 10.1063/1.882241, durr1992quantum2, Duerr:2003erc, durr2004quantum, passon2004isn, hiley2006delayed}. 
That approach, originally motivated by de Broglie's  analogy of the wave equation with beam optics\footnote{de Broglie's idea \cite{de1930introduction} was to interpret, before the invention of quantum mechanics, the newly found phenomena in analogy to wave optics. He envisaged a field of potentialities to guide the dynamics of particles that  later was identified with Schr\"{o}dinger's wave function. Bohm's objective \cite{Bohm:1951xw, Bohm:1951xx} was to overcome the Copenhagen interpretation of quantum mechanics and provide a model for the measurement process without artificially dividing the system into observed and observing subsystems, and thus avoid the obscure notion of the collapse of the wave function. For a discussion see also \cite{rosen1945waves, hiley2005non, valentini2010broglie}.}, is carried out in the polar or eikonal representation of the wave function, assigning the front of the wave function a guiding function for classical point particles.

\subsection{The pilot-wave interpretation}

 By Eq.~\eqref{consprobab} the q.m. probability density is conserved.
%
In analogy to the continuity equation in classical fluid dynamics, diminishing density in a region is associated with an outgoing flow. 
We thus expect that the q.m. probability flow will accelerate upon entering a low-probability areas in order to "escape" this volume element, decelerate when entering regions with probability highs, and propagate with roughly constant velocity in areas with uniform probability distribution. 

\medskip
Bohm postulated the existence of classical positions of particles as "hidden" variables (later called also \emph{beables} by Bell \cite{bell2004speakable}), with their trajectories determined from the \emph{guidance equation}\footnote{This again is analogous to the wave and geometrical optics: Light rays propagate in a direction $\nu$ perpendicular to the phase front of the light wave, $\nu = \nabla.S$. The eikonal is related to the index of refraction of the medium, $S^2 = n^2$.}    
\begin{equation*}
    \dot{q}_i(t) = \frac{1}{m} \nabla S(q_i(t)).
 \end{equation*}
Here the index $i$ runs across the entire configuration of an $N$-particle system, its spatial dimension $\times$ number of particles. 
The particles are "guided " by the modulus of the total wave function for which de Broglie coined the expression \emph{pilot wave} \cite{de1930introduction}.
Each trajectory is determined by its initial position $q_i^0 = q_i(t_0)$ with the statistical likelihood given by the (conserved) distribution  $\rho(q_i(t_0))$, and the wave function of the entire system. 
The identification of the statistical distribution of initial positions of the particles with the quantum mechanical probability density, \eref{def:prob}, is called the \emph{quantum equilibrium hypothesis}.\footnote{The physical importance of deviations from the equilibrium are discussed in~\cite{Valentini:2024gzy}.}  The q.m. probability continuity equation then ensures its validity for all times (\emph{equivariance} \cite{durr1992quantum2}).

\medskip
In order to calculate both the particle trajectories and the associated probability, the wave function of the system must be calculated upfront to serve as input -- the pilot wave -- for the guidance equation and its boundary conditions. 
Even though the temporal evolution of the particle trajectories is determined solely by the eikonal, it is the contribution of the quantum potential $V_{quant}$ that makes up the difference between classical and quantum mechanics.

\begin{figure*}[h]%
	\centering
	{{\includegraphics[width=\textwidth]{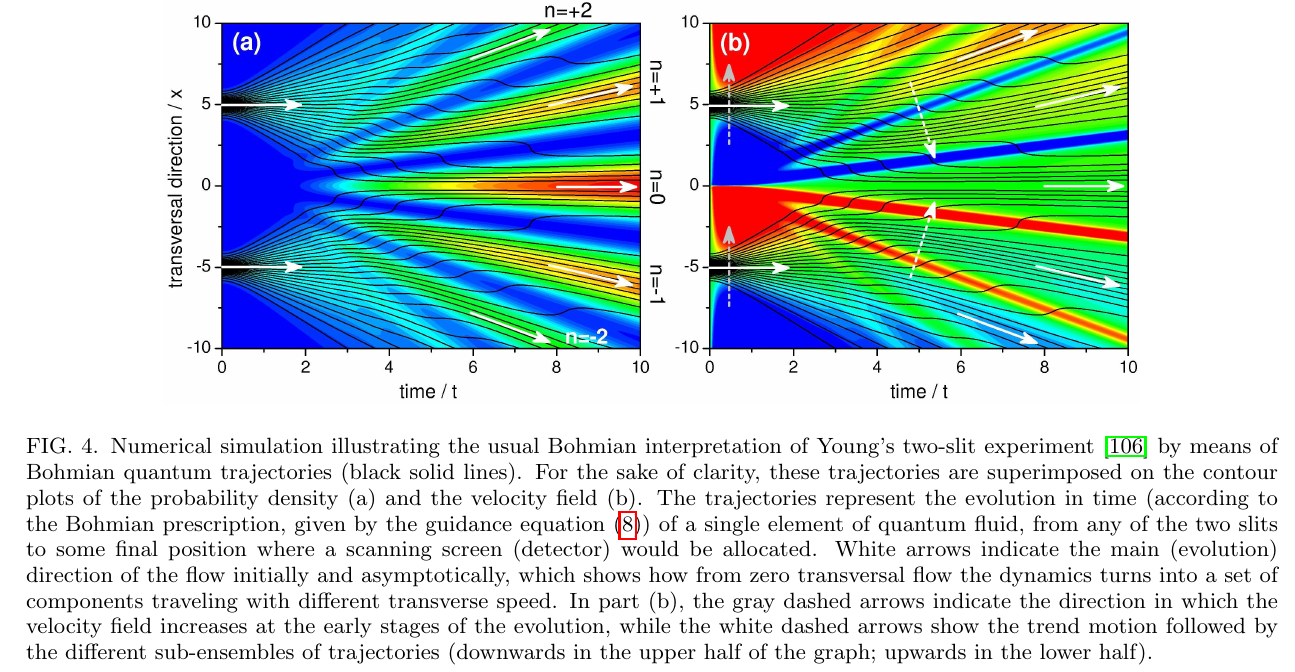} }}%
	\caption{\footnotesize Reproduced from Ref. \cite{sanz2019bohm}}%
	\label{fig:sanz}%
\end{figure*}

\medskip
As an illustrative example the classical double-slit experiment has been analyzed in \cite{philippidis1979quantum, Duerr:2003erc}  using Bohm's guidance equation. The "escape" behavior of the particle trajectories from low-probability regions has been visualized in Ref.  \cite{sanz2019bohm} -- and copied here in Fig. \ref{fig:sanz}. 

\medskip
We conclude:

\begin{itemize}
    \item 
    Bohm's interpretation appears as a comprehensive approach to quantum cosmology avoiding  the need for a macroscopic observer external to the observed system, and for  identification of measurements with operators, making the dilemma of operator ordering  obsolete. 
\item
    The uncertainty inherent to quantum theory is here of epistemic nature, ascribed to missing  knowledge about the initial condition of the true trajectory, rather than ontic as in the  orthodox Copenhagen framework \cite{von2013mathematische}. Alternative non-standard schemes for quantum mechanics (many-worlds theory, decoherence) appear less comprehensive as well \cite{10.1063/1.882241,chen2021quantum, Duerr:2003erc, everett1957relative}. 
\item
    The eikonal $S$ is implicitly, via the conservation of probability, influenced by all other "particle trajectories". The theory is thus fundamentally non-local, circumventing the problem addressed in the EPR paradoxon and the violation of Bell's inequality condition \cite{Bell:1964kc}. 
\end{itemize}

\medskip
With the pilot-wave approach to quantum cosmology the linear, first-order, guidance equation can be solved only after having first solved \eref{Friedmanqt} for the universal wave function, and subsequently  calculated  the eikonal $S_\mathrm{U}(a,\tau)$ and the probability density $\rho_\mathrm{U}(a,\tau)$.  
This requires in the first place fixing the boundary condition for the wave function, and identifying the effective wave function of the Universe in line with known observational facts. 

\subsection{The initial conditions for Bohmian trajectories}
The boundary conditions for the wave function at the origin shall be reasonably defined from mathematical and physical considerations. 

Combining now the framework of weak measurements, as laid out in~\sref{sec:EI}, and the equilibrium hypothesis of the de Broglie-Bohm framework, enables to fix the initial conditions of the particle trajectories for some $\tau = \tau_i$.
The statistical distribution of the starting point of the trajectory is given by the quantum mechanical probability density $\rho_\mathrm{U}(a,\tau;\bar{K},\Omega) = \left|\psi_\mathrm{U}(a,\tau;\bar{K},\Omega)\right|^2$.
The quantum guidance field, in addition, derives from the eikonal $S_\mathrm{U}(a,\tau;\bar{K},\Omega)$, via~\eref{def:eikonal}.

\medskip
In the following two options will be addres,sed for the initial condition determining the solutions of the first-order ordinary guidance equation 
\begin{equation} \label{cond:eikonal}
    \frac{\d q_{\bar{\mathrm{K}}}(\tau)}{\dtau} = \frac{1}{2} \pfrac{}{a} S_\mathrm{U}(a)|_{a=q_{\bar{\mathrm{K}}}(\tau)},
\end{equation}
that here describes the spatial expansion of the Universe.

\subsubsection{Inflation}

With the physical desire to ensure an inflationary scale expansion in the vicinity of the origin we explicitly request $\rho_U(0) = 0$. 
The \emph{spatial} distribution of the initial conditions in the de Broglie-Bohm approach shrinks to just one point.

\medskip
As seen at the example of the double-slit experiment,  any trajectory to rapidly escape the region a low q.m. probability around the origin, $\rho_\mathrm{U}(a \approx 0) \approx 0$, and eventually proceed along (smooth) high probability ridges. 
Thus requesting the wave function to be quenched from the origin appears instrumental for generating an accelerating expansion dynamics, i.e.\ inflation \cite{Pinto-Neto:2004szq}. 
The \emph{inflationary} initial condition is thus  
\begin{equation}
\frac{q_{\bar{\mathrm{K}}}(\tau)}{\dtau}|_{\tau = \tau_i} \rightarrow \infty,
\end{equation}
for some initial instant $\tau_i$, and we expect
\begin{equation}
q_{\bar{\mathrm{K}}}(\tau_i) \approx 0. 
\end{equation}

\subsubsection{The Hubble slot}



Rather than starting with the inflation / Big Bang scenario, we rely here on facts derived directly from observations at present-day, i.e.\ at $\tau = \tau_0$ and $q_0 \equiv q(\tau_0) = a_0$. 
Of course, when starting from this point to explore the past of the Universe, the calculation  must be directed backwards in time, from present day, $\tau_0$, to $q_i = q(\tau_i)$  whatever might be the size of the Universe at~$\tau_i$, the beginning of time. 

Setting $\tau_0 = 1/h$, as defined in \eref{def:Hubbletime}, for the present-day, the instant $\tau_i$, at which the Universe was born, need not be zero, it can even become negative. The age of the universe
\footnote{The popular assignment of the Hubble time $\tau_0(h) = \tau_{100}/h$ to the age of the Universe assumes an eternally constant expansion rate given by the Hubble constant $h H_{100}$.} 
is then $\tau_{tot} = \tau_{0} - \tau_i$.    

\medskip
With the above assumption the initial configuration of the trajectory in space and time is determined by the values of $h$ and $a_0$.

\medskip
The value of $a_0$ is found from a constraint based on physical observations: 
The slope of the trajectory $q(\tau_0)$ should give the present-day expansion velocity, $\d a(t)/\dt(t=1/H_0) = H_0$. In the units used here this gives
\begin{equation}
\frac{\d q_{\bar{\mathrm{K}}}(\tau)}{\dtau}|_{\tau = \tau_0} = h,
\end{equation}
implying the constraint 
\begin{equation} \label{eq:restrS}
    \onehalf \pfrac{}{a} S_\mathrm{U}(a=a_0,\tau=1/h;\Omega) = h
\end{equation}
for the eikonal $S_\mathrm{U}(\Omega)$. 
Here $\Omega$ stands for all physical parameters of the theory, providing some  freedom for aligning the theory with observations.

\medskip

For simplicity we set $a_0(\bar{h}) = 1$ for $\bar{h}$ being  the relative Hubble constant for vanishing FLRW curvature $\bar{K}$. 
(In astronomical studies $\bar{K}=0$ is considered as the most likely, \emph{Concordance}, value anyway.) 
We have to realize, though, that the scale $a_0$ denotes a scale factor \emph{relative} to some absolute dimension, say $r_0$, that depends on the parameters of the geometry, here $K$.  Hence $a_0\,r_0(K) \ne a_0\, r_0({K'})$. 
A natural measure for the absolute scale $r_K$ of the Universe is the so called event horizon $a_0(K)/H(K))$. 
If for simplicity the scale $a_0(\bar{K}) = 1$ is set for $\bar{K} = 0$, then
\begin{equation}
a_0(K) = a_0(\bar{K}) \frac{H(K)}{H({\bar{K})}} = \frac{h_K}{h_{\bar{K}}}  = \frac{h_K}{\bar{h}}.
\end{equation}

The initial configurations of the Bohm trajectories are thus restricted to an interval given by the empirical values of $h$. 
This is consistent with the restriction of the wave function to its effective form as discussed in Section~\ref{sec:EI}.
In analogy to restricting possible  particle trajectories in the classical double-slit experiment \cite{philippidis1979quantum} to the actual slit widths,  we dubbed it the "Hubble slot". 

\section{Summary and outlook} \label{sec:summary}
An alternative approach to Quantum Cosmology (QC) has been developed for a class of generic extended theories of gravity. 
In this approach the 3$^{rd}$ quantization is carried out a posteriori, i.e.\ only after translating the Friedman equation into a classical mechanics problem of a single non-relativistic particle moving in a potential.
Thereby the problem of missing time in QC is avoided.
The normalized cosmic time is shown to be conjugate to the spatial curvature parameter, $\Omega_\mathcal{K}$, which appears as the eigen-value of the quantized Hamiltonian. 
The wave function of the Universe emerges then as a superposition of corresponding eigen-solutions of that quantized Hamiltonian.
The respective boundary conditions are briefly addressed. 
By analyzing in detail the impact of (weak) measurements, i.e.\ astronomical observations, the expansion coefficients are fixed 
to restrict the wave function of the Universe to an  effective wave function. 
These coefficients are chosen to be Gaussians reflecting the measured Hubble rate and its error bars.
Facing the fact that an external observer is excluded in Quantum Cosmology, we invoke the de Broglie-Bohm interpretattion of quantum mechanics
and discuss the Bohmian guidance equation for the scale parameter of space.   
Thereby the boundary conditions ("Hubble slot") of the Bohmian trajectories are chosen to reproduce the current scale, $a = 1$, and the current expansion rate. 

A detailed study leading to numerical results will be the subject of a forthcoming paper.
With a concrete numerical study we not only envisage  to prove the viability of this approach, but also expect to win new insights on the different phases of the cosmic evolution, for example if pockets in more complex potentials from extended gravity give rise to tunneling effects and re-write the history of the Universe. . 
The steps lying in front of us are highly complex, though.
Calculating the potential from a specific extended theory of gravity, selecting  the boundary conditions for the wave function and for the guidance equation, determining the eigenfunctions and the wave function of the Universe, deriving  the  eikonal and solving the guidance equation constitute a demanding programme.




\thispagestyle{empty}

\end{document}